\documentclass[12pt]{article}
\usepackage{amssymb,amsmath,latexsym,amsfonts,amsmath}
\usepackage{cite}
\usepackage{graphicx}
\usepackage{comment}

\makeatletter
\@addtoreset{equation}{section}
\makeatother

\def\bra{\langle}
\def \ket{\rangle}
\def\Lam{\Lambda}
\begin{document}
%%%%%%%%%%%%%%%%%%%%%%%%%%%%%%%%%%%%%%%%%%%

\baselineskip 0.7cm

\begin{titlepage}

\begin{flushright}
{\bf KUNS-2330 }\\
\today
\end{flushright}

\vskip 1.35cm
\begin{center}
{\bf{Asymptotically Vanishing Cosmological Constant in the Multiverse}}

\vskip 1.2cm
Hikaru Kawai, Takashi Okada
\vskip 0.4cm

{{ Department of Physics, Kyoto university, Kyoto 606-8502, Japan}}\\

\vskip 1.5cm

\abstract{
We study the problem of the cosmological constant in the context of the multiverse in Lorentzian spacetime, and show that the cosmological constant will vanish in the future. 

This sort of argument was started from  Coleman in 1989, and he argued that the Euclidean wormholes make the multiverse partition a superposition of various values of the cosmological constant $\Lam$, which has a sharp peak at $\Lambda=0$. However, the implication of  the Euclidean analysis to our Lorentzian spacetime is unclear. With this motivation, we analyze the quantum state of the multiverse in Lorentzian spacetime by the WKB method, and calculate the density matrix of our universe by tracing out the other universes. Our result predicts vanishing  cosmological constant. While Coleman  obtained the enhancement at $\Lam=0$ through the action itself, in our  Lorentzian analysis  the similar enhancement arises from the front factor of $e^{iS}$ in the universe wave function, which is in the next leading order in the WKB approximation. }
\end{center}

\end{titlepage}

\section{Introduction and Conclusion}One of the major problems of particle physics  and cosmology is the smallness of the observed value of the vacuum energy, that is the cosmological constant $\Lam$. We must explain why $\Lam$ is many orders of magnitude smaller than the  Planck scale\cite{wormhole}. One of the most promising attempts to solve this problem is the one based on the Euclidean wormhole effect first proposed by Coleman \cite{coleman1988there}\footnote{In \cite{banks1988prolegomena}, Banks also discussed the effect of bi-local interaction. In this paper, we mainly follow Coleman's argument.}and studied closely by other authors\cite{cline1989does, strominger1989lorentzian, polchinski1989phase}\cite{giddings1989baby}. In this paper, we study this mechanism in the context of Lorentzian spacetime.

\begin{figure}[htbm]
\begin{center}
\includegraphics[width=3cm]{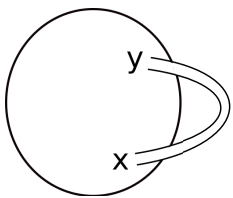}
\caption{A sketch of an euclidean universe with a small wormhole insertion. The thin tube represents the wormhole. }
\label{fig:euclidean_universe}
\end{center}
\end{figure}

As we review in Section $ \ref{sec:baby}$, a microscopic wormhole  effectively induces the following bi-local interactions  in the partition function of the Euclidean Einstein-Hilbert action,
\begin{equation}
\sum_{i,j}  \ c_{ij} e^{-2S_{\text{wh}}} \int d^4 x d^4 y \nonumber
\sqrt{g(x)}\sqrt{g(y)}
O_i(x)O_j(y),
\end{equation}
where $S_{\text{wh}}$ is the semi-wormhole action and $x,\ y$ represent the points into which the wormhole  is inserted(see Fig.\ref{fig:euclidean_universe}). 

Summing over any number of wormholes, it turns out that the system  is described by the following  effective action, 
\begin{equation*}
S_{\text{eff}}=S_E + e^{-S_{\text{wh}}}(a^\dagger+a)\int d^4 x\sqrt{g
(x)},
\end{equation*}
where we have introduced a pair of operators $a$ and $a^\dagger$, satisfying $[a,a^\dagger]=1$, which can be interpreted as a creation (annihilation) operator of a baby universe. 
The effect of $a$ and $a^\dagger$ is to make the cosmological constant dynamical, in the sense that it should be integrated as \footnote{In fact, there are many models in which the wave function of the universe naturally becomes a superposition of various possible values of the cosmological constant. See, for example,  \cite{unruh1989unimodular}\cite{ng1990possible}\cite{smolin2009quantization}\cite{Shaw:2010pq}\cite{hawking1984cosmological}\cite{bousso2000quantization}. In our argument, however, all universes in the multiverse should have the same cosmological constant, which follows naturally in the baby universe mechanism.}
\begin{equation*}
Z_{\text{universe}} = \int \mathcal{D}g d\Lam \exp\bigl({-\int d^4x \sqrt{g}(R-2 \Lam)} \bigr).
\end{equation*}

\begin{figure}[htbp]
\begin{center}
\includegraphics[width=5cm]{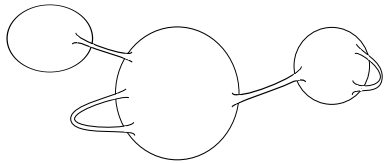}
\caption{A sketch of an example of the Euclidean multiverse. Parent universes are interacting through baby universes.}
\label{fig:euclidean_multiverse}
\end{center}
\end{figure}

Coleman analyzed which value of $\Lam$ is favored in the partition function. The result is that the integral over the metric can be approximated by a 4-sphere solution and leads to a $e^{\frac{1}{\Lam}}$ factor as follows,
\begin{equation*}
Z_{\text{universe}} \sim \int d\ {\Lam} \ e^{\frac{1}{\Lam}}.
\end{equation*}
If one include the effect of other parent universes, which are connected with ours through  baby universes (see Fig.\ref{fig:euclidean_multiverse}), then the multiverse partition function becomes $\exp(\exp(\frac{1}{\Lam}))$. 
From this argument, he claimed that the wormhole effect could solve the cosmological constant problem.\\

\begin{figure}[htbp]
\begin{center}
\includegraphics[width=7cm]{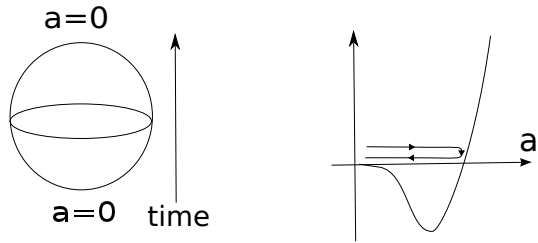}
\caption{The 4-sphere solution can be interpreted as a foliation of 3-spheres whose radius expands from zero to $\frac{1}{\sqrt{\Lam}}$ and then shrinks to zero.}
\label{fig:bounce}
\end{center}
\end{figure}

What does this result of the Euclidean gravity imply to  Lorentzian spacetime? Naively, the 4-sphere solution can be interpreted as a bounce solution. Therefore, the action associated with the solution is  expected to be the  amplitude of a universe tunneling form nothing to the size of the 4-sphere (see Fig.\ref{fig:bounce}). 
However, if we computes the tunneling amplitude directly by the WKB method as Vilenkin did\cite{vilenkin1984quantum}, we obtain a factor $e^{-\frac{1}{\Lam}}$, instead of $e^{\frac{1}{\Lam}}$ as Coleman did. Therefore, the physical meaning of the 4-sphere solution is unclear, and whether or not  the mechanism  can work  in Lorentzian gravity is  doubtful.

\begin{figure}
\begin{center}
\includegraphics[width=8cm]{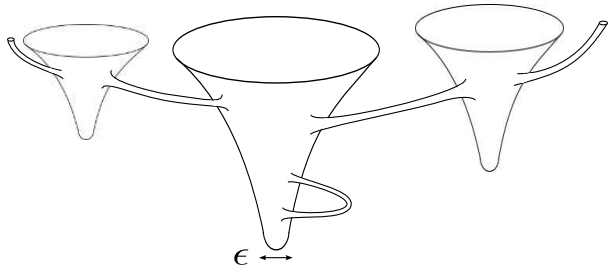}
\caption{A sketch of an example of the  multiverse. Parents universes emerges with a small size $\epsilon$ by a tunneling process. In this example, the initial state has no baby universes and the final state has two baby universes.}
\label{fig:multiverse}
\end{center}
\end{figure}

In this paper, we analyze the multiverse in the Lorentzian framework, and see that the cosmological constant problem is indeed resolved by the baby universes. First, in Section \ref{sec:baby}, we review the derivation of the effect of baby universes in the Euclidean path integral and obtain its Lorentzian counterpart via a Wick rotation (see Fig.\ref{fig:multiverse}, which is the Lorentzian version of Fig.\ref{fig:euclidean_multiverse}). 

In  Section \ref{sec:wavefunction}, assuming  that there are initially  no baby universes and that the parent universe starts from a small size $\epsilon$ via a tunneling process, we calculate the parent universe wave function $\phi_{E=0}(z)$, where $z$ is the size\footnote{Strictly speaking, $z \equiv  a^3/9$ has a dimension of volume. However, for the sake of simplicity, we call it ``size''.} of the universe, by using the WKB approximation. The result is 
\begin{eqnarray*}\phi_{E=0}(z) = \frac{1}{\sqrt{\pi/2}\sqrt{z}\sqrt{k_{E=0}(z)}}\sin (\int_{0}^z k_{E=0}(z')dz' +\alpha ),
\end{eqnarray*}
where $E=0$ represents the so-called Hamiltonian constraint, which we will discuss later. The factor $\frac{1}{\sqrt{k_{E=0}(z)}}$ behaves like $\frac{1}{\Lam^{1/4}}$ for large $z$ and plays an important role for our mechanism.

 From this, we can write down the N-universe wave function as follows,
\begin{equation*}
\Phi_N(z_1^{(N)},\cdots, z_N^{(N)})\equiv 
\int d \lambda \ \mu^{N}\phi_{E=0}(z_1^{(N)})\cdots\phi_{E=0}(z_N ^{(N)})\otimes \ 
e^{-\frac{\lambda^2}{4}} 
 |  \lambda\ket,
\end{equation*}
where $\lambda$ represents eigenvalues of $a+a^\dagger$ and is related to $\Lam$ as $\Lambda \equiv  \frac{1}{2}e^{-S_{\text{wh}}}\lambda  + \Lambda_0$, and $\mu$ is the probability amplitude associated with creation of an universe\footnote{In fact, there is also a numerical factor coming from the path integral calculation. We will define $\mu$ more precisely in Section \ref{sec:wavefunction}.}
.
The point is that the quantum state of  N-universe is a superposition of  various $\lambda$, or  various cosmological constants $\Lam$. Then the state of the multiverse is given by
\begin{equation*}
 |\phi_{\text{multi}} \ket \sim \sum_{N=0}^\infty  |\Phi_N \ket.
\end{equation*}
 
 In Section \ref{sec:densitymatrix}, we compute the density matrix of our universe by tracing out the other universes. 
The result is 
\begin{eqnarray*}
\rho(z,z') \sim \int_0^\infty d \Lambda \ e^{-\frac{\lambda^2}{2}} |\mu|^2 \phi_{E=0}(z')^* \phi_{E=0}(z) \exp(\frac{|\mu|^2}{\sqrt{9 \pi^2 \Lambda}}\times \log z_{IR} ),
\end{eqnarray*}
where $z_{IR}$ is an infrared cut-off of the size of the universes. The density matrix  has a strong peak around $\Lam=0$, which agrees with Coleman's assertion. However, our mechanism is essentially different: While the Euclidean analysis obtains the enhancement $e^{\frac{1}{\Lam}}$ through the action itself, we obtain the above enhancement factor  from the front factor $\frac{1}{\sqrt{k(z)}}$ of $e^{iS}$, which is in the next leading order in the WKB calculation. 

 In Section \ref{sec:discussion}, we explain that what our model predicts to vanish is not  the present value of the cosmological constant but that in the far future. We also discuss the generality of our mechanism.

%%%%%%%%%%%%%%%%%%%%%%%%%%%%%%%%%%%%%%%
\section{Effect of Baby Universes} \label{sec:baby}

We first review the Coleman's argument on the effect of baby universes \cite{coleman1988there}(see also \cite{klebanov1989wormholes}). We start from the Euclidean Einstein gravity with a bare cosmological constant  $\Lambda_0$,
\begin{equation}
\int\mathcal{D} g \exp( - S_E)=\int\mathcal{D} g_{\mu \nu} \exp(-\int d^4x \sqrt{g}(R- 2 \Lambda_0)). \nonumber
\end{equation}

A wormhole configuration effectively contributes to the partition function as follows,
\begin{equation}
\int \mathcal{D}g\ \frac{1}{2}  \sum_{i,\ j}\ c_{ij} e^{-2S_{\text{wh}}} \int d^4 x d^4 y
\sqrt{g(x)}\sqrt{g(y)}
O_i(x)O_j(y)  \exp(-S_E),
\end{equation}
where  $c_{ij} $  are some constants and $S_{\text{wh}}$ is the action of a semi-wormhole. Summing over the number of wormholes amounts to the factor

\begin{equation}
\exp\biggl(\frac{1}{2}  \sum_{i,\ j}\ c_{ij} e^{-2S_{\text{wh}}} \int d^4 x d^4 y \nonumber
\sqrt{g(x)}\sqrt{g(y)}
O_i(x)O_j(y) \biggr).
\end{equation}
If we restrict our attention in the case that  $O(x),O(y)$ are the identity operator $\hat{1}$ , the interaction  becomes
\begin{equation}
\exp\biggl(\frac{1}{2} e^{-2S_{\text{wh}}}  \int d^4 x d^4 y \ \sqrt{g(x)}\sqrt{g(y)}  \biggr),\label{eq:wormhole_effect}
\end{equation}
where we have absorbed the numerical constant $c_{ij}$ into $S_{\text{wh}}$.
By introducing an auxiliary variable $\lambda$, this bi-local interaction can be rewritten as a local interaction as follows,
\begin{eqnarray}
\int d \lambda \exp\bigl(-e^{-S_{\text{wh}}}\lambda \int d^4x \sqrt{g(x)} -\frac{\lambda^2}{2} \bigr),  \label{eq:integrationoverlambda}
\end{eqnarray}
where the first term in the exponent additively  contributes to the cosmological constant. Including wormhole configurations is, thus, 
equivalent to  summation over the cosmological constant.

Alternatively, we can express the  wormhole effect by  using  the following  Lagrangian
\begin{equation}
S_{\text{eff}}=S_E + e^{-S_{\text{wh}}}(a^\dagger+a)\int d^4 x\sqrt{g
(x)}.  \label{eq:creationannihilation}
\end{equation}
 Here, we have introduced a pair of operators  $a$ and  $\ a^\dagger$ satisfying $[a,a^\dagger]=1$, which can be interpreted as a creation/annihilation operator of a baby universe\footnote{To understand this formula, one considers an amplitude between the initial and final state both with no baby universe $\bra\Omega| \exp\bigl(e^{-S_{\text{wh}}} (a+a^\dagger) \int d^4x \sqrt{g}\bigr) |\Omega \ket $. 
By using the Baker-Campbell-Hausdorff formula, it is easy to show that this amplitude recovers  Eqn.(\ref{eq:wormhole_effect}).}.  The last term can be regarded as a part of the cosmological constant in each eigenspace of $a+a^\dagger$. Although  (\ref{eq:integrationoverlambda}) and  (\ref{eq:creationannihilation}) are equivalent,  (\ref{eq:creationannihilation})  is more convenient  to construct the wave function of the universe.

Finally, we obtain the Lorentzian counterpart by a Wick rotation,
\begin{equation}
S=\int d^4 x \sqrt{-g(x)}(R- 2 \Lambda_0)-e^{-S_{\text{wh}}}(a^\dagger+a)\int d^4 x\sqrt{-g(x)}. \label{eq:start}
\end{equation}
 We use this action to study the cosmological constant problem in the context of Lorentzian quantum cosmology.

\section{Quantum State of the Multiverse}\label{sec:wavefunction}
Equation (\ref{eq:start}) implies that the quantum state of the whole multiverse including baby universes can be expressed as a tensor product of the  parent universes' state and the state of the space spanned by  $a$ and $a^\dagger$. In this section, we first determine the wave function of a parent universe $\phi_{E=0}(z)$ with size $z$. Then we construct the quantum state of the multiverse.
 
 We will make some assumptions about the multiverse state in order to make the physical picture as concrete as possible. However, the conclusion we will obtain does not depend on  these assumptions,  as is discussed in Section \ref{sec:discussion}.

 \subsection{Wave Function of a Parent Universe}\label{eq:parent_wavefunction}

We first diagonalize $a+a^\dagger$. For each eigenspace of $a+a^\dagger$ with eigenvalue $\lambda$, the action  becomes 
\begin{equation*}
\int\mathcal{D} g_{\mu \nu}  \exp(i S_{\Lambda})=\int\mathcal{D} g_{\mu \nu}  \exp(\int d^4 x \sqrt{-g(x)}(R- 2 \Lambda)),
\end{equation*}
where $\Lambda \equiv  \frac{1}{2}e^{-S_{\text{wh}}}\lambda  + \Lambda_0$.

We will consider the homogeneous, isotropic and closed universe:
\begin{equation}
ds^2 = - N(t)^2 dt^2 + a(t)^2 d \Omega^2, \label{eq:metric}
\end{equation}
where $d \Omega^2$ is the metric on a unit 3-sphere.

\begin{figure}[htbp]
\begin{center}
\includegraphics[width=15cm]{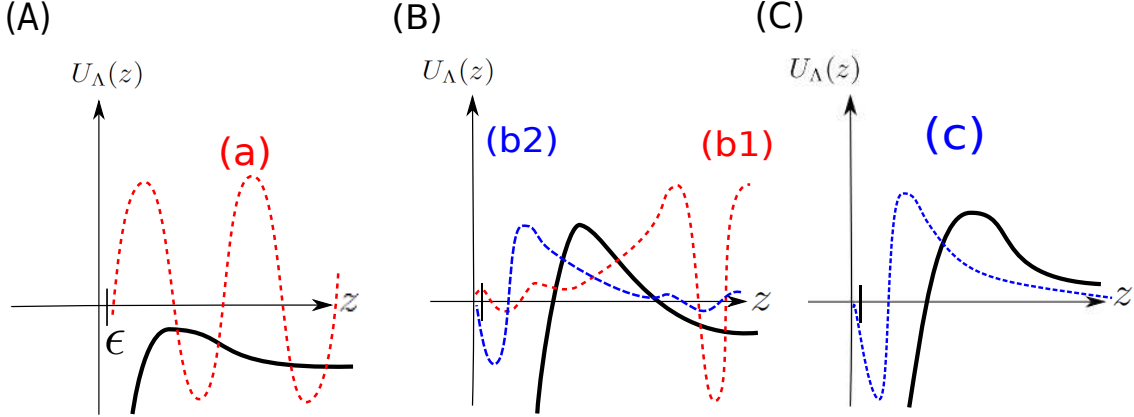}
\caption{Three types of potential (A)(B)(C) and the associated four wave functions(a)(b1)(b2)(c) are sketched. The solid line represents the potential $U_{\Lam}(z)$, and the colored dashed line represents the wave function of the universe $\phi_{E=0}(z)$. $Case$(A):$\Lam>0$ and the energy contribution from  matter and radiation is large. This corresponds to our universe.  $Case$(B):$\Lam>0$ and the energy contribution from  matter and radiation is small. Two wave functions (b1) and (b2) correspond to the two classical solutions. $Case$(C):$\Lam<0$.}
\label{fig:potential}
\end{center}
\end{figure}

Substituting the metric (\ref{eq:metric}), the action becomes 
\begin{eqnarray*}
S_{\Lambda} &=& -\frac{1}{2} \int dt \  N\bigl[a \dot{a}^2/N^2-(a-\Lambda a^3)\bigr].
\end{eqnarray*}
In terms of $z(t) := \frac{a(t)^3}{9}$, it can be expressed as
\begin{equation*}
S_\Lam= -\frac{1}{2} \int dt \  N\bigl[ \dot{z}^2/z N^2-\bigl((9z)^{1/3}- 9 \Lambda z \bigr)\bigr].
\end{equation*}
The momentum $p_z$ conjugate to $z$ is given by $p_z = -\dot{z}z/N$, and
the Lagrangian can be written in the canonical form,
  \begin{equation*}
L_\Lam = p_z \dot{z} - N \mathcal{H}_\Lam,
\end{equation*}
where
\begin{eqnarray}
\mathcal{H}_\Lambda(p_z,z) 
&:=& z[-\frac{1}{2}p_z^2- U_\Lam(z)] ,\ \  \text{where\  } U_\Lambda(z) :=\frac{9^{1/3}}{2 z^{2/3}}-\frac{9}{2}\Lambda  \label{eq:ham} .
\end{eqnarray}
If we take account of the existence of matter or radiation, the potential becomes
\begin{equation}
U_\Lambda(z) :=\frac{9^{1/3}}{2 z^{2/3}}-\frac{9}{2}\Lambda-\frac{C_{matt.}}{z}-\frac{C_{rad.}}{z^{4/3}},
\end{equation}
where $C_{matt.}$ and  $C_{rad.}$ are some constants associated with matter and radiation energies. The potential is classified into three types, and we will consider the associated four wave functions (see Fig.\ref{fig:potential}).

To quantize this system via path integral, we take the following metric on the configuration space 
\begin{equation}
|| \delta g_{\mu \nu }||^2= \int d^4 x \sqrt{-g}g_{\mu \nu} g_{\rho \lambda} \delta g^{\mu \rho}\delta g^{\nu \lambda} \propto \int dt ( \frac{a^3}{N} (\delta N)^2+ N a (\delta a)^2),
\end{equation}
which is invariant under general coordinate transformation, and this metric leads to the volume form of the functional integral
\begin{equation}
\underset{t}{\Pi}  \ a^2 \delta N \delta a  \propto \underset{t}{\Pi}  \delta N \delta z := [dN][dz].
\end{equation}
Collecting these results,  the path integral  becomes a quantum mechanical system,
\begin{eqnarray}
\int [dN][dz][dp_z] \exp(i \int dt (p_z \dot{z} - N \mathcal{H}_\Lambda)),
\end{eqnarray}
where $\mathcal{H}_\Lambda$ is defined by (\ref{eq:ham}).

\begin{figure}[bthm]
\begin{center}
\includegraphics[width=4cm]{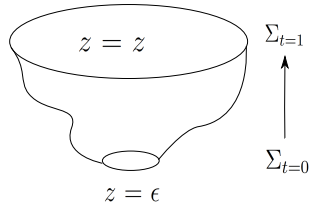}
\caption{The path integral (\ref{eq:pathintegral}) is defined as a sum over all histories connecting two geometries.}
\label{fig:history}
\end{center}
\end{figure}

In the rest of this section,
we will determine the quantum state of the universe, assuming that  the universe initially has a small size $\epsilon$(see Fig\ref{fig:history}), 

The amplitude between $z=\epsilon$ and $z=z$ is given by the following path integral,
\begin{eqnarray}
\bra z |e^{-i \hat{\text{H}}} |\ \epsilon \ket=\underset{z(0)=\epsilon,\ z(1)=z}{\int} [dp_z][dz][dN]\exp(i \int_{t=0}^{t=1}dt\  (p_z \dot{z} -N(t) \mathcal{H}_{\Lam})). \label{eq:pathintegral}
\end{eqnarray}
By choosing the gauge such that $N(t)$ is constant $\text{T}$,  the path integral of $N(t)$ is reduced to the ordinary integral over $-\infty<\text{T}<\infty$\footnote{To be precise, we should integrate only positive $T$ if we fix the time-ordering of the surface $\Sigma_{t=0}$ and $\Sigma_{t=1}$ as in Fig.\ref{fig:history}. However, we take the integration range as $-\infty<\text{T}<\infty$  to obtain the well-known Wheeler-Dewitt equation in the path integral formalism. This procedure corresponds to summing over the ordering of the two surfaces too.},
\begin{eqnarray*}
&&\int_{-\infty}^{\infty} d \text{T} \underset{z(0)=\epsilon,\ z(1)=z}{\int} [dp_z][dz]\exp\biggl(i \int_{t=0}^{t=1} dt\  (p_z \dot{z} -\text{T} \mathcal{H}_{\Lam})\biggr)\\
&=& \mathcal{C}\times \int_{-\infty}^{\infty}  d \text{T}\bra z |e^{-i {\text{T}}\mathcal{H}_\Lam} |\ \epsilon \ket \\
&=&\mathcal{C}\times  \bra z | \delta(\mathcal{H}_{\Lam})| \epsilon\ket \\
&=&\mathcal{C} \times  \bra z | \delta({\mathcal{H}_\Lam}) \biggl(\int_{-\infty}^{\infty} dE | \phi_{E}\ket \bra \phi_{E} |\biggr) | \epsilon\ket,
\end{eqnarray*}
where $\mathcal{C}$ is some constant and we have inserted  the complete set \{$|\phi_E\ket$ \}  defined by\footnote{Here, we have assumed that the spectrum of the universe is continuous although it does not seem true for the Type (b2) or (c) universe since these states are sort of bound states. The justification of this assumption is discussed in   \ref{app:scalar_field}.}
\begin{subequations}
\begin{eqnarray}
\bra \phi_E' |\phi_E \ket = \delta(E-E'), \label{eq:normalization}\\
 \mathcal{H}_{\Lam}|\phi_E\ket =E |\phi_E\ket .\label{eq:hamiltonianeq}
\end{eqnarray}
\end{subequations}
Therefore, by using $\phi_{E}(z) \equiv \bra z | \phi_E\ket$, the amplitude can be expressed as
\begin{eqnarray}
\mathcal{C}\times   \phi^*_{E=0}(\epsilon) \phi_{E=0}(z). \label{eq:amplitude}
\end{eqnarray}
In other words, the quantum state of the single universe that emerged with size $\epsilon$ is\footnote{This analysis is similar to that of \cite{cline1989does}.}
\begin{equation}
\mathcal{C}\times \phi^*_{E=0}(\epsilon)| \phi_{E=0}\ket.\label{eq:universestate}
\end{equation}

We can calculate $\phi_{E}(z)$  in the canonical quantization formalism. 
By replacing $p_z \rightarrow - i \partial / \partial  z$ in the Hamiltonian (\ref{eq:ham}), Eqn.(\ref{eq:hamiltonianeq}) becomes 
\begin{equation}
\sqrt{z}\bigl(\frac{1}{2} \frac{d^2 }{d z^2}-U_\Lam(z)\bigr) \sqrt{z} \ \phi_E(z)=E \phi_E(z)\label{W-D}.
\end{equation}
Note that this leads to the  Wheeler-DeWitt equation for $E=0$. However, we need to solve this equation for general $E$ in order to determine the normalization of the wavefunction. After short calculation, it is reduced to the following differential equation
\begin{equation}
(-\frac{d^2}{d z^2}-k^2_E(z))\sqrt{z}\phi_E(z) =0,
\end{equation}
where 
\begin{eqnarray*}
k_E^2(z) &\equiv& -2U_\Lam(z)-\frac{2E}{z} \\
&=&9\Lambda-\frac{2E}{z}
-\frac{9^{1/3}}{z^{2/3}}+ \frac{2C_{matt.}}{z}+\frac{2C_{rad.}}{z^{4/3}} \nonumber.
\end{eqnarray*}
By applying the WKB method to the function $\sqrt{z}\phi_E(z)$, we can easily solve this equation.

%%%%%%%%%%machigai
\begin{comment}
\begin{eqnarray}
&&(-\frac{d^2}{dz^2}- \frac{1}{z}\frac{d}{dz}- k_E^2(z) )\phi_E(z)=0, \label{eq:WKB}\\
 &&\text{\ where\ } k_E^2(z) \equiv -\frac{1}{4z^2}-\frac{9^{1/3}}{z^{2/3}}+9\Lambda+ \frac{2C_{matt.}}{z}+\frac{2C_{rad.}}{z^{4/3}}+\frac{2E}{z} \nonumber.
\end{eqnarray}

%%%%%%%%%%%%%%%%

%%%%%%%%%%% tsukawanai
We solve this equation by using the WKB method. 
Substituting $\phi_E(z) = e^{f(z)}$ into  Eqn.(\ref{eq:WKB}) leads to

\begin{equation}
f'' + f'^2 -\frac{1}{z}f' + k_E^2(z) =0.
\end{equation}
Then by writing $f= f_0 + f_1+ \cdots$ and collect the terms at each order, we obtain   the following equation
\begin{eqnarray}
{f'_0}^2&=& - k_E^2(z) \\
f'_1 &=& -\frac{1}{2} (\frac{f''_0}{f'_0}+\frac{1}{z})
\end{eqnarray}
%%%%%%%%%%%%%%
\end{comment}

Before that, we recall that there are four types of universe according to the matter and radiation energy and $\Lam$ (see the wave functions (a), (b1), (b2), (c) in Fig.\ref{fig:potential}.). In order to write down the solution, we must specify to which type of  the universe it belongs. As we will see in the next section,  the relevant quantity for the density matrix of our universe is the integral 
$\int d z |\phi^*_{E=0}(\epsilon) \phi_{E=0}(z)|^2$ (see  the exponent in the density matrix (\ref{eq:densitymat}), where $\mu$ is defined by ( \ref{eq:chemicalpotential})). For  Type (a),  (b1) and (b2), the integrals are all divergent.  As compared to Type (a), however, $\phi^*_{E=0}(\epsilon)$ is suppressed for Type (b1), and $\phi_{E=0}(z)$  for large $z$ is  small for Type (b2). Further, the integral for Type (c) is finite\footnote{As noted before, it is necessary to modify the model such that $E$ can take continuous values for Type (c). The detailed discussion is given in \ref{app:scalar_field}.}.  Therefore, we will concentrate on the universe of Type (a) as the first approximation.

 For Type (a), the whole region of $z>0$ is classically allowed and the general solution is given by a linear combination of 
\begin{eqnarray}
\phi_E(z) = \frac{1}{\sqrt{\pi}\sqrt{z}\sqrt{k_E(z)}}\exp(\pm i \int^z dz' k_E(z') ),\label{eq:general_solution}
\end{eqnarray}
where the normalization is determined from Eqn.(\ref{eq:normalization}) and the detailed calculation is given in  \ref{app:norm}.
To determine the solution completely, we must specify the boundary condition at $z=0$.  
As the simplest example, we require that $\phi_{E} (z=0)$ vanish.\footnote{The boundary condition might be more complex because the behavior in $z<\epsilon$ is highly related with the unknown dynamics of the singularity. In fact, our discussion is valid even if we choose other boundary conditions, which is discussed in Section \ref{sec:discussion}.}  The solution satisfying this condition is
\begin{eqnarray}
\phi_{E=0}(z) = \frac{1}{\sqrt{\pi/2}\sqrt{z}\sqrt{k_{E=0}(z)}}\sin (\int^z dz' k_{E=0}(z') ). \label{eq:positivelambdasolution}
\end{eqnarray}

\subsection{Wave Function of the Multiverse} \label{sec:multiverse}
We can construct the wave function of the whole multiverse from the single universe wave function $\phi_{E=0}(z)$. First, we recall that the quantum system of the multiverse has a pair of $a$ and  $a^\dagger$, and the eigenvalue $\lambda$ of $a+a^\dagger$ plays the same role as the cosmological constant. Parent universes in a fixed-$\lambda$ sector feel  the net cosmological constant $\Lambda =  \frac{1}{2}e^{-S_{\text{wh}}}\lambda  + \Lambda_0$, and evolve according to it.  Second, to determine the multiverse state, we must specify the initial distribution of $\lambda$. For example, if there are initially no baby universe  
as in Fig. $\ref{fig:multiverse}$, then the initial quantum state 
  can be written as  $  \int  d \lambda \ e^{-\lambda^2 /4} |\lambda \ket := |\Omega \ket$\footnote{It might be helpful to regard $a+a^\dagger$ as the position operator $\sqrt{2}x$ of a harmonic oscillator. }. We will take this case for concreteness although it is not relevant for our argument.
 
 In order to construct the quantum state of the multiverse $|\phi_{\text{multi}} \ket$, we need the probability amplitude of a universe emerging from nothing to  the size $\epsilon$, which we denote as $\mu_0$. Together with the factor in Eqn.(\ref{eq:amplitude}) or (\ref{eq:universestate}), the weight of each universe $\mu$ is given by,
\begin{equation} \mu :=\mu_0\times \mathcal{C} \times \phi^*_{E=0}(\epsilon). \label{eq:chemicalpotential}
\end{equation}
A crucial fact is that $\mu $ does not depend on $\Lam$ strongly. This is because  $\phi^*_{E=0}(\epsilon)$ is a smooth function of $\Lambda$ as is seen from (\ref{eq:positivelambdasolution}), and $\mathcal{C}$ arising from the path measure should have nothing to do with $\Lam$.
 
  Collecting these pieces together, we find
\begin{equation}
|\phi_{\text{multi}} \ket = \sum_{N=0}^\infty  |\Phi_N \ket,
\end{equation} 
where $|\Phi_N \ket$ stands for the N-universe state, whose wave function is given by
\begin{equation}
\Phi_N(z_1^{(N)},\cdots, z_N^{(N)})= 
\int d \lambda \ \mu^{N} \phi_{E=0}(z_1^{(N)})\cdots\phi_{E=0}(z_N ^{(N)})\otimes \ 
e^{-\frac{\lambda^2}{4}} 
 |  \lambda\ket. \label{eq:multiverse_state}
\end{equation}
Here, we note that $\mu$ and $\phi_{E=0}(z_i)$ depend on $\lambda$ via the net cosmological constant $\Lambda=  \frac{1}{2}e^{-S_{\text{wh}}}\lambda  + \Lambda_0$.

\section{Density Matrix of Our Universe} \label{sec:densitymatrix}
In this  section, we study the density matrix of our universe and analyze the preferred value of the cosmological constant. We will find that the density matrix has a strong peak around $\Lam =0$.

Our universe (denoted by $z\text{\ or}\  z'$)  can be regarded as a subsystem of the multiverse. Therefore, we can obtain the density matrix of our universe by tracing out the other universes (denoted by $z^{(N)}_i\text{\ or}\  z'^{(N)}_i, \ i=1 \cdots N$),
\begin{eqnarray}
\rho(z',z) &=& \sum_{N=0}^{\infty} \int \frac{dz^{(N)}_1 \cdots dz^{(N)}_N}{N!} \underset{\lambda}{\text{Tr}} \  \Phi_{N+1}(z', z_1^{(N)},\cdots, z_N^{(N)})^*\Phi_{N+1}(z, z_1^{(N)},\cdots, z_N^{(N)}) \nonumber\\
&=& \sum_{N=0}^{\infty} \frac{1}{N!} \int d \lambda \  e^{-\frac{\lambda^2}{2}} |\mu|^2\phi_{E=0}(z')^* \phi_{E=0}(z) \biggl( \int dz^{''} |\mu \phi_{E=0}(z^{''})|^2 \biggr)^N \nonumber\\
&\propto& \int d\Lam\   e^{-\frac{\lambda^2}{2}}|\mu|^2\ \phi_{E=0}(z')^* \phi_{E=0}(z) \exp\biggl( \int dz^{''} |\mu \phi_{E=0}(z^{''})|^2 \biggr). \label{eq:densitymat}
\end{eqnarray}
In the last line, we have changed the integration variable from $\lambda$ to $\Lam$ and  the factor $e^\frac{-\lambda^2}{2}$ is understood as a function of $\Lam$. 

\begin{comment}
As we have seen in Section \ref{eq:parent_wavefunction}, we can justify the assumption in the previous section that the dominant contribution comes from the Type (a) rather than (b) or (c), since for Type (b)or(c) the integral of the wave function is suppressed compared with Type (a). 
we  evaluate the integral over $z$ in the exponent by using the result of the previous section.
As described before, there are three types of the universe. From Eqn.{\ref{eq:densitymat}},  we find that the dominant contribution comes from (a)-Type since the integral $ \int dz\phi^*_{E=0}(\epsilon)\phi_{E=0}(z^{''})$  is small for  (b)or(c)-Type  compared with Type (a)  because of the tunneling suppression. 
We  can thus assume all parent universes are (a)-Type as the first approximation, and 
\end{comment}

The integral  of the wave function can be evaluated by using the solution (\ref{eq:positivelambdasolution}) for Type (a) as
\begin{equation}
\int dz |\phi_{E=0}(z)|^2 = \int dz \frac{2}{\pi z k_{E=0}(z)} \sin^2( \int^z k_{E=0}(z')dz' )).
\end{equation}
 Actually, this integral is logarithmically divergent since  $k_{E=0}(z) \sim \sqrt{9\Lam}$ is constant for large $z$. If we regulate this divergence by introducing the upper bound $z_{IR}$ of the integration, which corresponds to the infrared cut-off of the size of universes, then the integral becomes 
\begin{equation}
\int^{z_{IR}} dz |\phi_{E=0}(z)|^2 \sim \frac{1}{\sqrt{9\pi^2\Lam}} \log z_{IR} + \cdots.
\end{equation}
Thus, the density matrix becomes
\begin{eqnarray}
\rho \sim \int_0^\infty d \Lambda \  |\mu|^2 e^{-\frac{\lambda^2}{2}} \phi_{E=0}(z')^* \phi_{E=0}(z) \exp(\frac{|\mu|^2}{\sqrt{9\pi^2\Lambda}}\times \log z_{IR} ). \label{eq:result_densitymat}
\end{eqnarray}
As we have discussed in  Section \ref{sec:multiverse}, $\mu$ does not have a strong $\Lam$-dependence. Therefore, the exponential function in (\ref{eq:result_densitymat}) provides a sharp peak at $\Lam=0$. This enhancement  comes from the front factor $\frac{1}{\sqrt{k(z)}}$ of $e^{iS}$ in the WKB approximation, which  is essentially different from Coleman's original argument, where the enhancement comes from the action itself.

\section{Discussion}\label{sec:discussion}
First, we point out an interesting feature of our model. According to the standard model of cosmology, the cosmological constant is measured to be a positive and small non-zero value.   Our model, on the other hand, seems to predict that the cosmological constant is exactly zero because of the enhancement factor $e^{\frac{1}{\sqrt{\Lambda}}}$ in the density matrix. However, the $\Lambda$ appearing in the enhancement factor is defined by the asymptotic value of $k_{E=0}(z) \rightarrow \sqrt{9\Lam}$ for $z \rightarrow \infty$. Therefore,  it is not  the presently observed  cosmological constant but its value in the far future that our model predicts to vanish. Thus, our model does not conflict with the current observation.

Second, we make some comments on the assumptions we made in Section {\ref{sec:wavefunction}}, and explain that our mechanism  works in  a quite generic condition. The solution in the classical region is generally a linear combination of
\begin{equation}
\frac{1}{\sqrt{2\pi}{\sqrt{z}}\sqrt{k(z)}} \exp( \pm i \int^z d z' k(z')),
\end{equation}
and to be concrete, we have chosen the sine-type solution (see Eqn.(\ref{eq:positivelambdasolution})).
If we consider another situation where the universe keeps expanding\cite{vilenkin1984quantum},  we should use the following solution\footnote{ The sign in the exponent can be understood from the relation $p_z =- z \dot{z}/N$.} \begin{equation}
\frac{1}{\sqrt{2\pi}{\sqrt{z}}\sqrt{k(z)}} \exp( - i \int^z d z' k(z')).
\end{equation}
However, the enhancement still occurs because it comes from the front factor $\frac{1}{\sqrt{k(z)}}$, which does not depend on the boundary condition.

\begin{figure}
\begin{center}
\includegraphics[width=7cm]{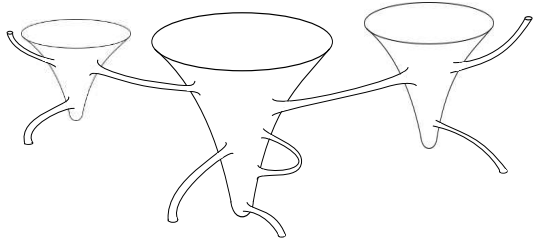}
\caption{A sketch of an example of the  multiverse. In this case, the initial state has some baby universes.}
\label{fig:multiverse_initialbabies}
\end{center}
\end{figure}

Another generalization is about the initial state of the baby universes.  In Section \ref{sec:wavefunction}, we assumed that there are initially no baby universes in the multiverse,  which is described by $\int d \lambda e^{-\frac{\lambda^2}{4}}|\lambda \ket$. However, our argument is valid even in the case where there are many baby universes at the beginning(see Fig.\ref{fig:multiverse_initialbabies}). 
If we consider a general superposition $\int d \lambda  \  f(\lambda) | \lambda \ket$ as the initial multiverse state, then the density matrix becomes 
\begin{eqnarray}
  \int d \Lambda \ f^2(\lambda=\lambda(\Lam))|\mu|^2 \phi_{E=0}(z')^* \phi_{E=0}(z) \exp(\frac{|\mu|^2}{\sqrt{9 \pi^2\Lambda}}\times \log z_{IR}),
\end{eqnarray}
which again has the peak at $\Lambda=0$ unless the function $f(\lambda=\lambda(\Lam))$ behaves singularly around $\Lam=0$.

From these considerations,  it seems that the mechanism we discussed in the paper  is highly generic, and the problem of the cosmological constant is naturally solved in  the context of the multiverse in Lorentzian spacetime.

\section*{Acknowledgement} 
The authors thank T. Azeyanagi for valuable comments. H.K also thanks Henry Tye for fruitful discussions. This work is supported by the Grant-in-Aid for the Global COE program "The Next Generation of Physics, Spun from Universality and Emergence" from the MEXT.

\appendix
\def\thesection{Appendix \Alph{section}}
\section{ Normalization of the Wave Function  }\label{app:norm}
\def\thesection{ \Alph{section}}

In this appendix, we check the wave function (\ref{eq:general_solution}) satisfies the normalization (\ref{eq:normalization}),
\begin{eqnarray*}
\int^\infty_0 dz \phi^*_{E'} (z) \phi_E (z) = \delta(E-E').
\end{eqnarray*}
Substituting the wave function, the left hand side is
\begin{eqnarray}
\int^\infty_0 dz \frac{1}{\pi z \sqrt{k_E(z)k_{E'}(z)}}\exp( \pm i \int^z dz' ( k_{E'}(z')-k_{E}(z'))). \label{eq:norm_appendix}
\end{eqnarray}
Note that the delta function can arise from the integral over the asymptotic region $z \rightarrow \infty$.
For large $z$, $k_E(z)k_{E'}(z)\sim {9\Lam}$ and $k_{E'}-k_{E}\simeq\frac{\partial k_E}{\partial E}(E'-E)\sim\frac{1}{\sqrt{9\Lam}z}(E'-E)$, where we have used $k^2_E \sim 9\Lam +\frac{2E}{z}+\cdots$.
 From these, we can check (\ref{eq:norm_appendix}) indeed gives,
\begin{eqnarray*}
\int^\infty d (\log z)\frac{1}{\pi \sqrt{9\Lam}}\exp({\pm i \frac{1}{\sqrt{9\Lam}}(E'-E)\log z })=\delta(E'-E).
\end{eqnarray*}

\def\thesection{Appendix \Alph{section}}
\section{ Continuous Spectra of Type (b2) and (c) Universe }\label{app:scalar_field}
\def\thesection{ \Alph{section}}

 In Section $\ref{sec:wavefunction}$, we assumed that the quantum state of the universe  takes a continuous spectrum.  In this appendix,  by  considering a model with a homogeneous scalar field $\eta(t)$  as illustration, we justify that the spectrum of  Type (b2) or (c)  can be made continuous, and the wave function of the universe  can be  normalized as in Eqn.(\ref{eq:normalization}). 
  
Introduction of  a spatially homogeneous scalar field $\eta(t)$ with the potential $V(\eta)$ gives the following  contribution to the original Hamiltonian $\mathcal{H}_\Lam$,
\begin{eqnarray}
\Delta \mathcal{H}_\Lam = z[\frac{p_\eta^2}{18z^2}+\frac{9}{2}V(\eta)]\nonumber, 
\end{eqnarray}
where $p_\eta$ is the canonical momenta given by  $p_\eta = \frac{\partial L}{\partial \dot{\eta}}= \frac{9z \dot{\phi}}{N}$. Here we will take $V(\eta)=0$.  

To show that the eigenvalue equation of  Type (b2) or (c)
\begin{eqnarray}
(\mathcal{H}+\Delta \mathcal{H}_\Lam )\phi_{E}(z, \eta)= E \phi_{E}(z,\eta) \label{eq:eigeneq}
\end{eqnarray}
has a solution for any energy $E$, we write the wave function as $e^{i K \eta } f(z)$ with some real number $K$.
Then, it is easy to understand that (\ref{eq:eigeneq}) always has a solution by adjusting the constant $K$ appropriately depending on  the energy, which we denote by $K=K_E$.

Furthermore, for  Type (c),  because the wavefunction $f(z)$ is normalizable due to the tunneling suppression(see Fig\ref{fig:potential}), 
we have
\begin{equation}
\int dz  d \eta\  \phi^*_{E'}(z, \eta)_E\phi (z, \eta) =  \int dz  d \eta\  e^{i(K_E-K_{E'})\eta} f^*_{E'}(z) f_{E} (z)=\text{some finite value}\times \delta(E-E'),
\end{equation} 
where the delta-function is obtained by integrating over $\eta$ since $\delta(K_{E'}-K_E) \propto \delta(E'-E)$.
Thus, we can always make the solution $\phi(z,\eta)$ satisfy the normalization condition ({\ref{eq:normalization}}) by normalizing $f(z)$ appropriately.

\bibliographystyle{unsrt}
\bibliography{bunken}

\begin{thebibliography}{10}

\bibitem{wormhole}
\text{For reviews, see, for example, S. Weinberg. The cosmological constant
  problem.} {\it{rev, mod.phys, 61(1), 1989}}; {\text{s.m carroll. the
  cosmological constant.}}{\it{ living reviews in relativity, 4:1, 2001.}}

\bibitem{coleman1988there}
S.~Coleman.
\newblock {Why there is nothing rather than something: A theory of the
  cosmological constant* 1}.
\newblock {\em Nuclear Physics B}, 310(3-4):643--668, 1988.

\bibitem{banks1988prolegomena}
T.~Banks.
\newblock {Prolegomena to a theory of bifurcating universes: a nonlocal
  solution to the cosmological constant problem or little lambda goes back to
  the future}.
\newblock {\em Nuclear Physics B}, 309(3):493--512, 1988.

\bibitem{cline1989does}
J.M. Cline.
\newblock {Does the wormhole mechanism for vanishing cosmological constant work
  in lorentzian gravity?}
\newblock {\em Physics Letters B}, 224(1-2):53--57, 1989.

\bibitem{strominger1989lorentzian}
A.~Strominger.
\newblock {A Lorentzian analysis of the cosmological constant problem}.
\newblock {\em Nuclear Physics B}, 319(3):722--732, 1989.

\bibitem{polchinski1989phase}
J.~Polchinski.
\newblock {The phase of the sum over spheres}.
\newblock {\em Physics Letters B}, 219(2-3):251--257, 1989.

\bibitem{giddings1989baby}
S.B. Giddings and A.~Strominger.
\newblock {Baby universe, third quantization and the cosmological constant}.
\newblock {\em Nuclear Physics B}, 321(2):481--508, 1989.

\bibitem{unruh1989unimodular}
WG~Unruh.
\newblock {Unimodular theory of canonical quantum gravity}.
\newblock {\em Physical Review D}, 40(4):1048, 1989.

\bibitem{ng1990possible}
Y.J. Ng and H.~van Dam.
\newblock {Possible solution to the cosmological-constant problem}.
\newblock {\em Physical review letters}, 65(16):1972--1974, 1990.

\bibitem{smolin2009quantization}
L.~Smolin.
\newblock {Quantization of unimodular gravity and the cosmological constant
  problems}.
\newblock {\em Physical Review D}, 80(8):084003, 2009.

\bibitem{Shaw:2010pq}
Douglas~J. Shaw and John~D. Barrow.
\newblock {A Testable Solution of the Cosmological Constant and Coincidence
  Problems}.
\newblock {\em Phys. Rev.}, D83:043518, 2011.

\bibitem{hawking1984cosmological}
SW~Hawking.
\newblock {The cosmological constant is probably zero}.
\newblock {\em Physics Letters B}, 134(6):403--404, 1984.

\bibitem{bousso2000quantization}
R.~Bousso and J.~Polchinski.
\newblock {Quantization of four-form fluxes and dynamical neutralization of the
  cosmological constant}.
\newblock {\em Journal of High Energy Physics}, 2000:006, 2000.

\bibitem{vilenkin1984quantum}
A.~Vilenkin.
\newblock {Quantum creation of universes}.
\newblock {\em Phys. Rev. D;(United States)}, 30(2), 1984.

\bibitem{klebanov1989wormholes}
I.~Klebanov, L.~Susskind, and T.~Banks.
\newblock {Wormholes and the cosmological constant* 1}.
\newblock {\em Nuclear Physics B}, 317(3):665--692, 1989.

\end{thebibliography}

\end{document}